# Brain Connectomes Come of Age

**Xiao-Jing Wang**[1,4], **Ulises Pereira**[1], **Marcello G. P. Rosa**[2], **Henry Kennedy**[3]

[1] Center for Neural Science, New York University, 4 Washington Place, New York, NY 10003, USA

[2] Biomedicine Discovery Institute and Australian Research Council Centre of Excellence for Integrative Brain Function, Monash University, Clayton, VIC 3800, Australia

[3] Stem Cell and Brain Research Institute, INSERM U846, 69500 Bron, France

[4] Correspondence: xjwang@nyu.edu

**Abstract.** Databases of directed- and weighted- connectivity for mouse, macaque and marmoset monkeys, have recently become available and begun to be used to build dynamical models. A hierarchical organization can be defined based on laminar patterns of cortical connections, possibly improved by thalamocortical projections. A large-scale model of the macaque cortex endowed with a laminar structure accounts for layer-dependent and frequency-modulated interplays between bottom-up and top-down processes. Signal propagation in a version of the model with spiking neurons displays a threshold of stimulus amplitude for the activity to gain access to the prefrontal cortex, reminiscent of the ignition phenomenon associated with conscious perception. These two examples illustrate how connectomics may inform theory leading to discoveries. Computational modeling raises open questions for future empirical research, in a back-and-forth collaboration of experimentalists and theorists.

- Directed- and weighted inter-areal cortical connectivity matrices of macaque, marmoset and mouse exhibit similarities as well as marked differences.
- The new connectomic data provide a structural basis for dynamical modeling multi-regional cortical circuit and understanding the global cortex.
- Quantification of cortical hierarchy guides investigations of interplay between bottom-up and top-down information processes.

## Introduction

In 1991 Felleman and van Essen published a landmark work of neuroanatomy that combined data from the existing literature to establish a hierarchy of the macaque monkey cortex [1]. This paper provided an impetus for efforts that, 10 years later, led to an inter-areal cortical connectivity matrix, the Collation of Connectivity data on the Macaque brain (CoCoMac) [2]. The CoCoMac matrix was fairly rough, with connections between area pairs assigned as absent, weak or strong (hence not quantitatively weighted); many connections were missing because of lack of information. Nevertheless, it represented a pioneering event in the field now referred to as brain connectomics.

The past two decades have seen significant advances [3–9]. Novel technologies have made it possible to determine wiring of neural circuits in the brain on microscopic, mesoscopic and macroscopic spatial scales [10–13]. Importantly, while it may be true that a picture is worth a thousand words, systematic measurements translated into precise numbers are essential for discovering general principles of large-scale cortical organization. This short review covers recent advances in our description of the cortico-cortical connections, and computational modeling based on the new quantitative databases. We shall summarize recent approaches and findings, as well as challenges that need to be addressed in order for the field to move ahead. The word "connectome" is currently used to refer to collations of data obtained with different methods, with multiple resolutions. The present review focuses on the connectome defined with cellular-resolution tracers, which at present can only be used in nonhuman animals.

## From anatomy to multi-regional cortical dynamics

In recent years new databases of inter-areal connectivity have become available for both macaque monkey [14-15, 16••, 17], mice [18–19, 21••], and marmoset [22, 23••]. Using a systematic analysis of retrograde tracing, the weight of cortico-cortical connection is indexed between 0 and 1 (FLNs, or fraction of labeled neurons), which measures the weight of connection from a source area relative to all source areas for a target area [15]. Therefore, connections are weighted parametrically, which is much more informative than a binary matrix. It is also directed, unlike diffusion tractography which, although non-invasive, cannot differentiate fibers from area A to area B and those in the reversed direction. Whereas fiber amounts can be inferred from tractography, measurements from tract tracing are direct and thus constitute a "ground truth". Indeed, the correlation is modest between log-transformed tractography and tracer connection weights in the macaque ($r \simeq 0.59$ [24]).

Three results are noteworthy. First, the Felleman-van Essen cortical hierarchy is significantly improved by quantification, so that a directed connection from area A to area B is assigned an FLN value, and areas are arranged along a one-dimensional hierarchy numerically. Second, the weight of connection (if present) between two areas decays exponentially with their distance (the exponential distance rule) [17]. Third, the weights of inter-areal connections are highly heterogeneous, spanning five orders of magnitude [15]. These three salient characteristics have also been shown in marmoset, another monkey species of growing interest in Neuroscience [22–23]. Therefore, a graph-theoretical view of cortical networks is inadequate unless spatial relationships between areas are taken into consideration [25]. This finding inspired a new class of generative models for the cortical networks that are explicitly spatially embedded [26], [14].

The new macaque connectivity matrix provides a structural scaffolding for the development of a dynamical model of multi-regional macaque cortex. As a sound practice in computational neuroscience, one must judiciously choose the level of complexity of a model that is proper to investigate specific scientific questions. In this case, the main question was: what is the biological mechanism that endows each area with appropriate temporal dynamics for its specialized function, such as rapid responses in the primary sensory areas and slow ramping activity underlying time integration in association areas? In the model, each area was mathematically modeled by a generic excitatory-inhibitory network, in accordance with the commonly accepted notion of a canonical circuit in the cortex [27–28]. The quantitative connection strengths, however, vary from one area to another. These variations were not random, but systematically change along low-dimensional axes across the cortical mantle. Chaudhuri et al. [29] considered the number of spines (loci of excitatory synapses) in the basal dendritic tree of pyramidal neurons, as a proxy of the strength of synaptic excitation per neuron, which displays an increasing gradient along the cortical hierarchy [30]. Interestingly, in this model, temporal dynamics of each area is dominated by a time constant that ranges from tens of milliseconds for early sensory areas to more than a second for prefrontal areas at the top of the cortical hierarchy, exactly what is desirable for functional differentiation. Importantly, the prevalent time constant of an area is not a monotonic function of its hierarchical position. For instance, the frontal eye field is at a relatively low position in the hierarchy [16], but it shows a long time constant by virtue of being part of the frontal lobe in close interactions with other frontal areas that display slow dynamics. The timescale spectrum in the cortex is constrained by both the macroscopic gradient of synaptic connection strength and the weighted inter-areal cortical network.

The new concept of macroscopic gradients [31••] applies to both synaptic excitation and inhibition. For instance, counts of diverse inhibitory cell-types across the mouse cortex revealed that the density of GABAergic cells expressing calcium-binding protein parvalbumin (PV), which control spiking outputs of excitatory pyramidal neurons, is the highest in the primary visual cortex and much lower in association areas [32–33]. Assessment of such macroscopic gradients can be carried out using a variety of data, including levels of gene expression that encode receptors for synaptic excitation and inhibition [34••, 35•] and neuronal density [36]. This approach allows identification of the biological fingerprint of different cortical areas; these data can then be incorporated into dynamical computational modeling.

They also are valuable for comparison across species. In particular, we will discuss below the definition of cortical hierarchy in primates versus rodents.

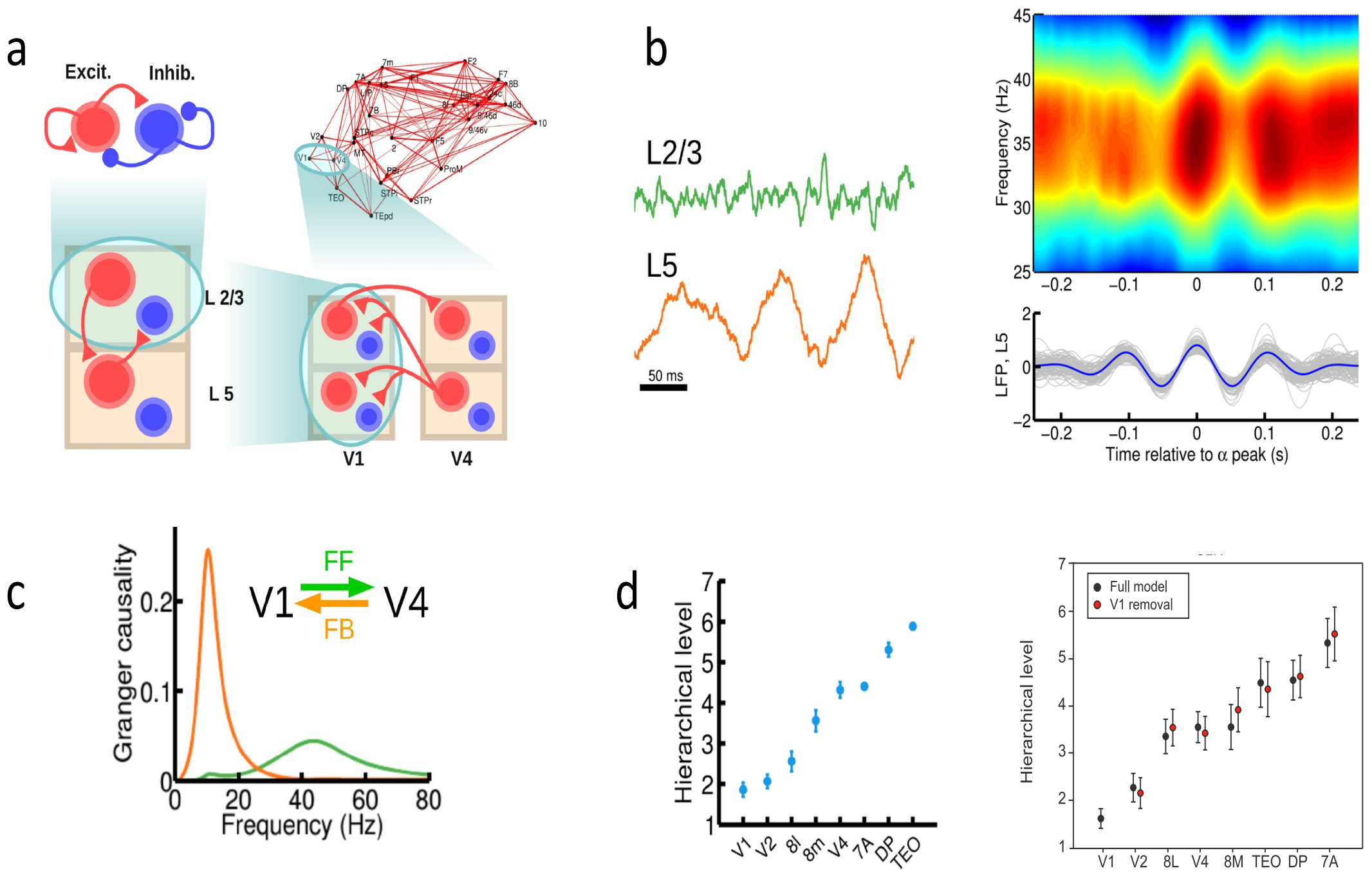

**Figure 1**: A multi-regional model of the macaque monkey cortex endowed with a laminar structure. (a) The scheme shows the four levels considered: a within-layer local microcircuit consisting of an excitatory(red) and an inhibitory (blue) population (upper left), a laminar circuit with two laminar modules (corresponding to supra- and infra-granular layers, lower left), an inter-areal circuit with laminar-specific projections (lower right), and a large-scale network of 30 cortical areas based on macaque anatomical connectivity (upper right). Each level is anatomically constrained. Only the connections at each level not shown at a lower level are plotted, for clarity. (b) Left panel: the superficial layer and deep layer display gamma (upper panel) and alpha (lower panel) oscillations. Right panel: the periodogram of the superficial layer shows gamma modulated by alpha wave (top), whereas the deep layer is dominated by alpha rhythmicity (bottom). (c) Granger causality as a function of frequency for feedforward signaling from V1 to V4 (green) and feedback (orange). (d) Cortical hierarchy deduced from the frequency-dependent Granger causality measure in the model (left panel) and in a monkey experiment (right panel). Reproduced with permission from [37] with experimental data from [76].

## Hierarchical information processing in the macaque cortex

Quantification of a cortical hierarchy is based on the observation that, in general, source neurons for a feedforward projection (e.g., from V1 to V2) reside in the superficial layers (above layer 4), whereas a feedback projection (V2 to V1) originates from neurons in the deep layers (below layer 4). Certain areas lack a prominent layer 4; some connections such as between V4 and FEF have similar proportions of source neurons in the superficial layers. Notwithstanding exceptions, it is clear that in order to investigate how bottom-up and top-down processes interact, a computational model should incorporate a laminar cortical structure. Mejias et al. [37] built such a model in which a local area has a superficial layer and a deep layer; each with an excitatory-inhibitory microcircuit (Fig. 1a). The superficial layer exhibits noisy synchronous oscillations in the gamma ($\simeq$ 40 Hz) frequency range [38], [39]; whereas the deep layer shows coherent oscillations at low beta ($\simeq$ 15-20 Hz) or alpha ($\simeq$ 10 Hz) frequency range [38], [40], [41]. The inter-laminar connections were calibrated based on the exiting literature. Consequently, gamma activities in the superficial layer were shown to be modulated by alpha (Fig. 1b), agreeing with empirical findings [42].

To assess the plausibility of this model, Mejias et al. [37] evaluated frequency-dependent Granger causality in a spontaneously active state. Granger causality is a measure of directionality of information flow. Monkey physiological studies showed that Granger causality is enhanced in the gamma frequency band for a feedforward projection (for example, from V1 to V4) but in the alpha frequency band for a feedback projection (V4 to V1) [43], [44]. This observation was captured by the model (Fig. 1c). Bastos et al. [44] found that the difference in the Granger causality peak values at the gamma and alpha frequencies could be used to establish a functional cortical hierarchy. The hierarchy, thus deduced purely by physiological measurements, is strongly correlated with that from the anatomical analysis in the macaque monkey. The large-scale laminar network [37] reproduces this hierarchy (Fig. 1d), thus substantially validating the computational model.

This model highlighted a few open questions that deserve attention in future experiments. In particular, inter-areal connection weights measured anatomically did not directly map onto physiological strengths of local synaptic connections, although they did show the same lognormal distribution [15], [45]. In the local microcircuit, synaptic strengths typically vary over 2-3 orders of magnitude [45] rather than five found in the inter-areal network [15]. Physiological data of this kind are currently not available for long-distance cortical projections. Moreover, the targets of a top-down projection are poorly understood. Existing forms of retrograde tracing [15] or anterograde fluorescent labeling of axons [18], [21] are not adequate; for example, labeled axonal terminals may target pyramidal cells in layers 2/3 and 5 with their distal dendrites in layer 1. Major distinct inhibitory neuron types have relative proportions that vary from area to area. They are differentially targeted by long-range connections, but this information is lacking at the present time. New methods are needed for quantification of projections with target specificity in terms of layers and cell types.

Modeling has also been used to re-visit a classical problem in computational neuroscience, namely, signal propagation across multiple neural populations [46•]. The problem is clearly formulated in a purely feedforward network, where neural group 1 receives an input and fires a burst of spikes, activating neural group 2, which in turn projects to neural group 3, etc [47]. Under different conditions, the signal either succeeds in propagating throughout the system, or it dies out in the middle of the network. The macaque cortical connectivity is endowed with ample, highly heterogenous, feedback projections. As a consequence, it is nontrivial to ensure reliable yet stable propagation of activity, say triggered by brief visual input to area V1. In a spiking neuron version of the multi-regional large-scale macaque cortex model, it was found that whereas activity increases with stimulus intensity in areas of the occipital lobe, those in the prefrontal cortex (PFC) exhibit near zero response when the stimulus intensity is below a threshold (Fig. 2a, upper panel, black versus red) [46]. In other words, the sensory stimulus needs to be sufficiently strong in order to be propagated along the cortical hierarchy and gain access to the PFC. This was not put by design but emerged unexpectedly in the model. Threshold crossing for access to the PFC has been hypothesized as a signature of awareness of a sensory input. When a stimulus appears in the environment with a small amplitude, we sometimes detect it, sometimes not. With the same physical stimulation onto our sense organs, the evoked activity remains largely confined to the posterior part of the cortex, and the input is reported as absent. When we are conscious of its presence, the Global Workspace Theory posits that the cortical core, largely centered in the PFC [14] “lights up" as in an “ignition" and activates the whole brain via feedback projections [48–50]. Further work is warranted to see if our model can indeed account for salient observations from monkey physiology about the ignition phenomenon [51], [52].

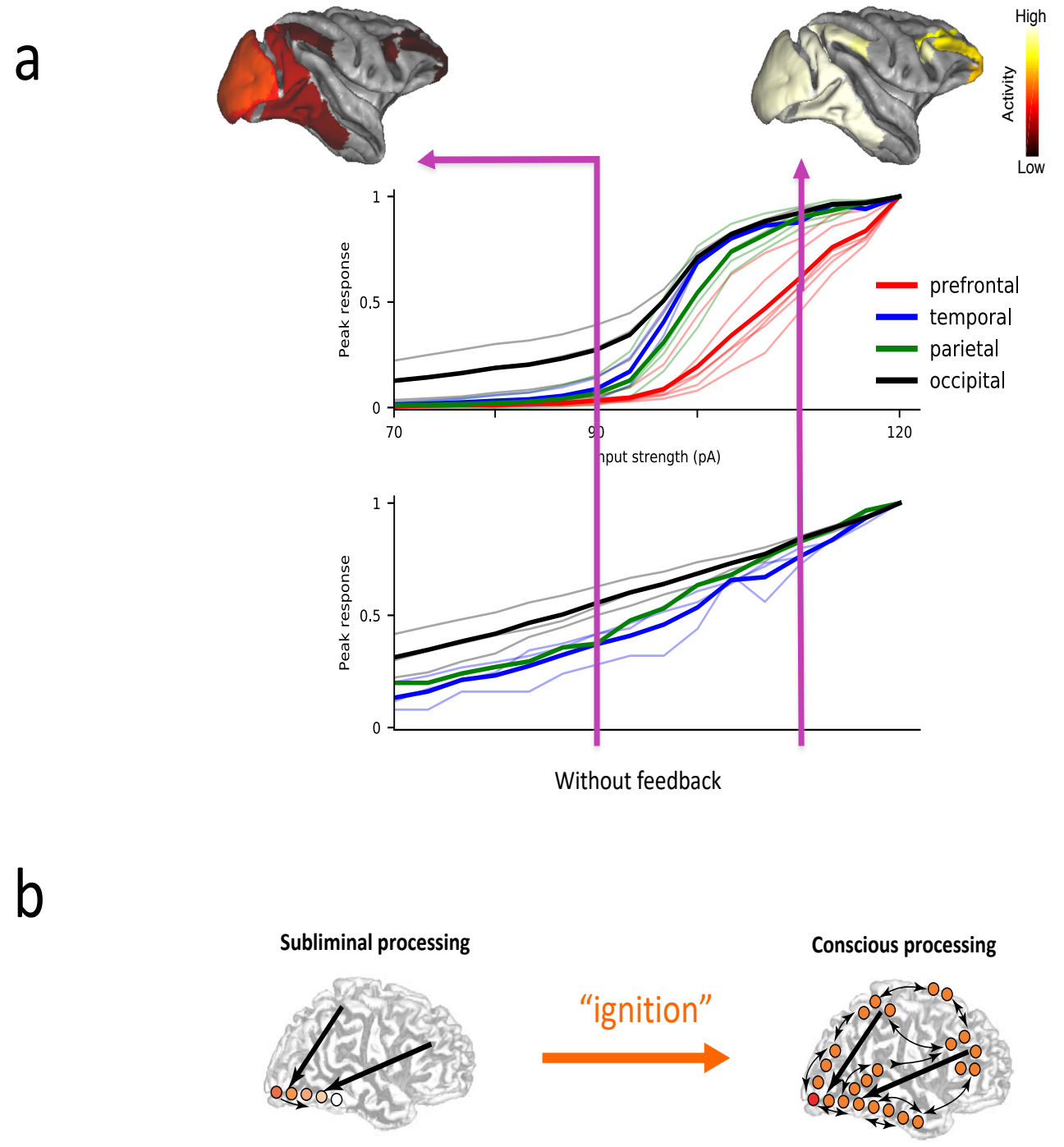

**Figure 2**: Signal propagation and the ignition phenomenon in the cortex. (a) Top and middle: In a macaque cortex model of spiking neurons, as the amplitude of an input to V1 is gradually increased, the peak response in areas of the occipital lobe (black) grows gradually. By contrast, activity is absent in the prefrontal cortex unless the stimulus intensity exceeds a threshold level (red). The activity map is confined to the posterior part of the cortex when the input is weak; if the input is above the threshold, access to the PFC leads to enhanced activity throughout the cortex. Note that this model included only a subset of cortical areas for which connectivity data are available, therefore the activity map is restricted only to those areas in the model. Bottom: The thresholding effect disappears when top-down connections in the model are deleted, demonstrating an important role of long-range feedback loops. (b) The model behavior is akin to the all-or-none ignition phenomenon associated with consciousness, that was observed experimentally with humans. Panel (a) is reproduced from [46], (b) from [49].

## Cortical hierarchy in mouse and marmoset

Cortico-cortical connectivity in mouse also displays a wide range of connection weights and the exponential distance rule [20, 53]. However, whether the mouse cortex displays a well-defined hierarchy remains unsettled. Previous studies note various biological entities are high in V1 and low in association areas, such as PV neuron density [32] and the T1w:T2w ratio from structural magnetic resonance imaging, which correlates with the level of myelin content in the grey matter [35, 54]. Such measures thus roughly change across the cortex in a way reminiscent of a hierarchy. However, many biomarkers exhibit statistical macroscopic gradients [31]. Ideally, one would like to identify an objective and robust definition of hierarchy, then assess the variations of circuit properties as dependent variables along the hierarchy defined as an independent variable.

A recent work examined the issue of hierarchy in the mouse cortex based on anterograde fluorescent labeling of axons [21••]. Using multiple Cre driver lines, Harris et al selectively traced layer- and cell type-specific projection patterns. An unsupervised method was used to consistently assign these laminar projections at the target area to be feedforward or feedback in a hierarchy. For instance, different types of thalamocortical connections targeting L4 versus L1 in the cortex were separately quantified. Interestingly, it was found that the inclusion of the thalamocortical projections enhanced the consistency of thus defined hierarchy [21••]. In a neurophysiological experiment using a mouse performing a detection task, the latency of spiking response to a visual stimulus was extracted from neurons in 6 visual areas [55], Person correlation of response latency with the anatomically defined Harris hierarchy was found to be quite high, about 0.9. A combination of anterograde fluorescent labeling and retrograde track tracing [56] promises to further establish a cortical hierarchy in the mouse.

Another approach was inspired by a recent study of the organization of transmodal default-mode networks in human and macaque [57•]. The work was based on a nonlinear dimensionality reduction method called diffusion map [58]. Briefly, the connectivity matrix is used to define an abstract *diffusion distance* between areas based on the transition probabilities of a hypothetical diffusion process. This distance produces a *diffusion space* where closer areas in this space share a larger number of paths connecting them, while areas far apart are less connected. In general, the diffusion distance depends on a low number of 'principal directions' or 'principal gradient' in diffusion space, providing the method a low dimensional embedding of the connectivity. Applying this approach to the whole mouse brain data of [59] and by choosing V1 as the origin in the diffusion space, a hierarchy among areas can be built by sorting areas by their diffusion distance to the origin.

Figure 3 shows the pairwise correlations between the Harris hierarchy, the hierarchy deduced from the diffusion map, PV density and T1w:T2w ratio in the mouse brain. Intriguingly, Spearman correlation coefficient values are in the range of 0.35 to 0.5. The explanation of substantial but far from perfect correlations is presently unclear, indicating that future research is warranted to achieve a consensus on the definition of cortical hierarchy in the mouse.

On the other hand, it is possible that a cortical hierarchy is flatter or less developed in rodents than primates [20]. This difference in organization could emerge from simple scaling laws [60-61], which predict that brain size is inversely correlated with "percent connectedness" (the fraction of brain cells with which any cell communicates directly). This, in turn, could have the effect of increasing the variety of inputs to any given cortical area, hence reducing the dominance of any single source, and "blurring" the definition of hierarchical levels.

Recent analyses based on a dataset of directed and weighted connections in the marmoset cortex shed light on this issue [22-23]. Marmosets, like macaques, are simian primates, but are much smaller (on average, the mass of the marmoset brain is 12 times smaller than that of *M. fascicularis*). In line with the scaling hypothesis, previous studies have indicated that the sources of afferents to both sensory [60] and association [61] areas are more widely distributed spatially across the cortex in marmosets than in macaques. However, a recent comprehensive study of the cortical connectome using statistical techniques applied to retrograde tracer data also revealed that this is accomplished without loss of specificity: the cortical connectivity matrix is very similar to that in the macaque in terms of overall density (approximately 2/3 of the possible connections that could exist are observed experimentally in both species), but they both differ from the mouse (where 97% of the possible connections exist). The similarity between macaque and marmoset extends to more detailed properties of the connectome, such as occurrence of reciprocal versus unidirectional connections. Other properties of the marmoset connectome, such as the presence of a well-defined core-periphery arrangement and the log-normal distribution of connection weights, also bring the two primates in close alignment. Importantly for the present argument, the marmoset cortex is also characterized by a well-defined hierarchy, where areas belonging to the different sensory domains occupy defined levels, from primary visual, auditory and somatosensory areas, though several higher-order association areas, to sensory association and polysensory areas. These multiple hierarchies converge to a core of frontal, posterior parietal, rostral temporal areas, which occupy the highest hierarchical levels, and include the regions of the cortex that expanded most clearly during primate evolution [63]. Furthermore, the hierarchical levels defined by connectivity are highly correlated with structural measures such as neuronal density and number of spines in the basal dendritic trees of pyramidal cells [64], [36]. Thus, the current evidence suggests that brain mass (and hence the number of neurons) does not in isolation fully predict the characteristics of the hierarchical organization of the cortex across mammals, and point to specific differences between primates and rodents, which are likely to have emerged due to specific evolutionary pressures. Further studies in marmosets, including the integration of cellular connectivity data with high-resolution tractography and functional connectivity measures using neuroinformatic platforms [65-66] offer the promise of greater insight onto the correlation with non-invasive measurements in the human brain, which promise to increase our ability to investigate the bases of neuropsychiatric conditions [67-69].

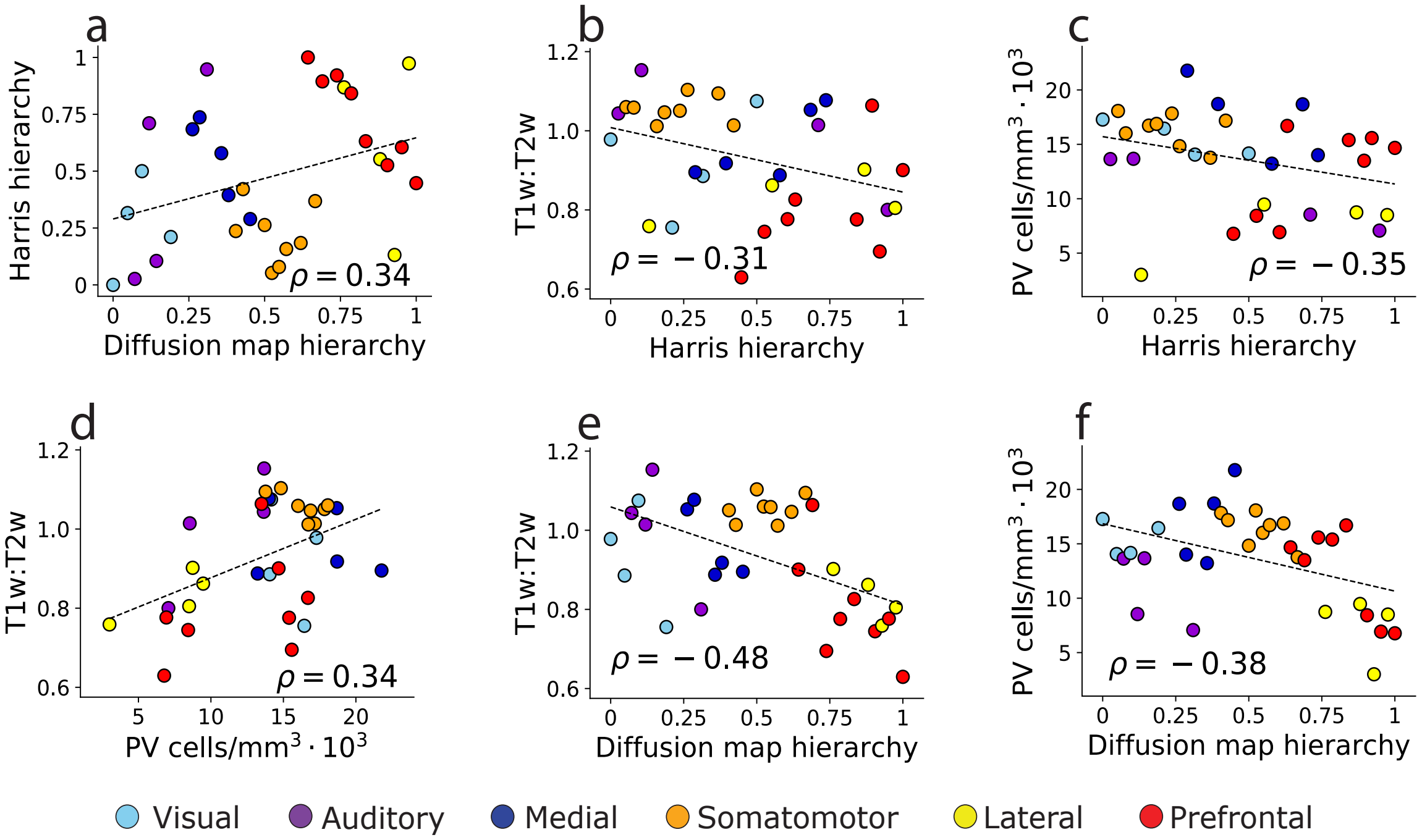

**Figure 3**: Cortical hierarchy in the mouse can be defined by four different measures: T1w:T2w ratio and PV neuron density decrease, whereas the diffusion map measure and the Harris hierarchy based on layer-dependent connectivity generally increase with the hierarchy. The pairwise correlation between these four measures, however, typically have a correlation of about 0.3-0.5.

## Looking into Future

In summary, directed- and weighted- inter-areal cortical connection matrices now exist for macaque, marmoset and mouse. These data are of a different kind from connectomics on μm spatial scale, achieved using electron microscopy for much smaller animals such as *Drosophila* fly [70]. Combined with genetic tools, research in this direction blurs the boundary between macroscopic and mesoscopic connectomes towards cell-type specific connectivity.

For monkeys, existent datasets are incomplete as they only include a subset of cortical areas compared to tractography, which provides a complete cortico-cortical connectivity matrix. This limitation makes it difficult for dynamical models to simulate functional connectivity, defined by the covariance of activities between cortical areas. A subnetwork does not encompass all areas and their feedback loops, and this could impact on global brain dynamics. While modeling has been attempted in this direction [71], it was done using tractography data, which has a poor signal-to-noise ratio (that is, it includes numerous false positives and false negatives) and is devoid of directionality information. Therefore, ongoing efforts to complete the full graph of monkey cortico-cortical connectivity should be a priority for the field.

The cortico-cortical connectivity discussed above is not cell-type specific. Empowered by genetic tools, future work will yield cell-type specific connection patterns, such as how different populations of pyramidal neurons in deep layers 5 and 6 project to distinct cortical and subcortical structures. Again, quantification into numbers would be required for such data to be utilized in computational modeling, which plays an increasing role in our investigations of complex cortical circuitry with its abundance of feedback loops.

A long list of open questions can be made for future research; here is a short one. First, how can we endow individual cortical areas with distinct information representations for different sensory modalities that comprehensively allow sensory coding to rule-guided decision making? Second, why do different circuits operate in different dynamical regimes, such as brief response, sequential activity and persistent activity? Third, can we harness genomic data to quantify biological properties in different cell types across cortical areas, that will inform future development of dynamical modeling? Fourth, what are the concise rules for the interactions between the cortex, the hippocampus, thalamus, amygdala, claustrum, basal ganglia and cerebellum?

In summary, technological advances, experiments and computational modeling have identified several general principles of large-scale cortical organization: (1) weights of inter-areal connections obey the exponential distance rule, (2) distributions of cortico-cortical connection weights are lognormal, (3) a cortical hierarchy can be parametrically quantified, (4) synaptic excitation and inhibition vary across the cortex in the form of macroscopic gradients. In the next phase of the brain connectome, genetically powered and cell-type specific connectome, single-cell RNA-seq mapping, large-scale neurophysiology using Neuropixels probes [72, 73•, 74••, 75] will produce a deluge of data. Novel analysis tools, new ideas and theory will be critical for us to transform data into knowledge, ushering in an era of computational neuroscience of the whole brain.

**Acknowledgments**. This work was partly supported by the ONR Grant N00014-17-1-2041, US National Institutes of Health (NIH) grant 062349, and the Simons Collaboration on the Global Brain program grant 543057SPI to XJW; by research grants from the Australian Research Council (DP140101968, CE140100007) to MGPR. UP was supported by The Swartz Foundation. HK was supported by LABEX CORTEX (ANR-11-LABX-0042) of Université de Lyon (ANR-11-IDEX-0007) operated by the French National Research Agency (ANR) (HK.), ANR-11-BSV4-501, CORE-NETS (HK.), ANR-14-CE13-0033, ARCHI-CORE (HK.), ANR-15-CE32-0016, CORNET (HK.), Chinese Academy of Sciences President's International Fellowship Initiative. Grant No. 2018VBA0011 (HK).